\documentclass{article}
\usepackage{spconf,amsmath,graphicx}

\usepackage{xcolor}
\usepackage{acronym}
\usepackage{amssymb}
\usepackage{multirow}
\usepackage{booktabs}
\usepackage{caption}
\usepackage{subcaption}

\acrodef{asc}[ASC]{Audio Scene Classification}
\acrodef{ts}[TS]{Temporal Support}
\acrodef{ir}[IR]{Instrument Recognition}
\acrodef{ssl}[SSL]{Self Supervised Learning}
\acrodef{sl}[SL]{Supervised Learning}
\acrodef{cnn}[CNN]{Convolution Neural Network}
\acrodef{ast}[AST]{Audio Spectrogram Transformer}
\acrodef{sld}[SLD]{Strongly Labeled Dataset}
\acrodef{wld}[WLD]{Weakly Labeled Dataset}
\acrodef{map}[mAP]{Mean Average Precision}
\acrodef{asc}[ASC]{Audio Scene Classification}
\acrodef{esc}[ESC]{Environmental Sound Classification}

\newcommand{\X}{\mathbf{X}}
\newcommand{\E}{\mathbf{E}}
\newcommand{\e}{\mathbf{e}}
\newcommand{\y}{\mathbf{y}}
\newcommand{\yh}{\hat{\mathbf{y}}}

\newcommand{\Yh}{\hat{\mathbf{Y}}}
\newcommand{\f}{f(\cdot)}
\newcommand{\g}{g(\cdot)}
\newcommand{\ts}{\delta_{t}}

\newcommand{\ie}{\textit{i.e.}\ }

\newcommand{\um}{\mu_{\text{m}}(\cdot)}
\newcommand{\ua}{\mu_{\text{a}}(\cdot)}

\title{On the choice of the optimal temporal support for audio classification with Pre-trained embeddings}
\name{Aurian Quelennec, Michel Olvera, Geoffroy Peeters, Slim Essid}
\address{LTCI - Telécom Paris, Institut Polytechnique de Paris}
\begin{document}
\ninept
\maketitle

© 2024 IEEE. Personal use of this material is permitted. Permission from IEEE must be obtained for all other uses, in any current or future media, including reprinting/republishing this material for advertising or promotional purposes, creating new collective works, for resale or redistribution to servers or lists, or reuse of any copyrighted component of this work in other works.

\begin{abstract}

Current state-of-the-art audio analysis systems rely on pre-trained embedding models, often used off-the-shelf as (frozen) feature extractors. 
Choosing the best one for a set of tasks is the subject of many recent publications. 
However, one aspect often overlooked in these works is the influence of the duration of audio input considered to extract an embedding, which we refer to as Temporal Support (TS). 
In this work, we study the influence of the TS for well-established or emerging pre-trained embeddings, chosen to represent different types of architectures and learning paradigms. 
We conduct this evaluation using both musical instrument and environmental sound datasets, namely OpenMIC, TAU Urban Acoustic Scenes 2020 Mobile, and ESC-50. 
We especially highlight that Audio Spectrogram Transformer-based systems (PaSST and BEATs) remain effective with smaller TS, which therefore allows for a drastic reduction in memory and computational cost. Moreover, we show that by choosing the optimal TS we reach competitive results across all tasks. In particular, we improve the state-of-the-art results on OpenMIC, using BEATs and PaSST without any fine-tuning.

\end{abstract}

\section{Introduction}
\label{sec:intro}

In the audio processing domain, the lack of vast task-specific annotated datasets has been a persistent challenge.
This scarcity has led to the development of models that are pre-trained on large-scale and diverse audio datasets, enabling them to generate very effective representations \cite{gemmeke2017audioset, Hsu21HUBERT}. 
These pre-trained models are subsequently either fine-tuned or used as frozen feature extractors when applied to smaller task-specific datasets \cite{Zaiem23finetune, angulo2023cosmopolite, Gururani19}. 

Notably,  pre-trained audio models are typically trained using either \ac{sl} \cite{hershey2017vggish, kong2020panns, Koutini22passt} or \ac{ssl} techniques \cite{Niizumi2023byola, Chen22beats, Cramer19openl3, Saeed21Cola}.
While \ac{cnn} architectures were once the prevailing choice, the newer \ac{ast} architectures \cite{Gong21ast} have proven to be more effective than \ac{cnn}s in numerous scenarios.
To assess the effectiveness and the generalization ability of such models across diverse downstream tasks, various benchmarking frameworks and challenges have been proposed \cite{turian2022hear, lWang2022hares, Yang2021superb, Niizumi2023byola}, aiding in the selection of the most suitable frozen pre-trained model to develop audio analysis systems.

However, one aspect that has not been systematically addressed in model evaluation is the impact of the duration of audio input considered for the extraction of an embedding, which we refer to as \ac{ts}. 
Remarkably, the subject of \ac{ts} has received limited consideration in current research, with only one study, to our knowledge, addressing it \cite{koutini21hear}. 
In this study, the authors show that altering the \ac{ts} of a PaSST model from $160$~ms to  $640$~ms improves the results on the DCASE 2016 Task 2 dataset \cite{Mesaros18task2}. 
Shedding light on the influence of the \ac{ts} in model performance, we conduct an in-depth evaluation of a selection of well-established and current state-of-the-art pre-trained models.

We selected the models to study based on four criteria: i) data used for training: we chose models that were pre-trained using AudioSet \cite{gemmeke2017audioset}, a dataset rich in both musical content and environmental sounds; ii) training paradigm: we considered \ac{sl} vs \ac{ssl} approaches; iii) type or architecture: \ac{cnn} vs \ac{ast} variants; and iv)  levels of complexity: choosing both lightweight and computationally intensive models. This resulted in the selection of BYOL-A \cite{Niizumi2023byola}, PaSST \cite{Koutini22passt}, and BEATs \cite{Chen22beats}. Needless to say, the selection was also motivated by the recent success of these models in various audio classification tasks.

Our evaluation of these frozen pre-trained audio embeddings extends across three distinct audio classification tasks: \ac{ir} using OpenMIC \cite{Humphrey18openmic}, \ac{asc} employing TAU Urban Acoustic Scenes 2020 Mobile \cite{Mesaros2018taudata}, and \ac{esc} utilizing ESC-50 \cite{Piczak15esc50}. 
We chose these datasets as reference results exist for them in previous works with most of the selected embeddings \cite{koutini21hear, lWang2022hares, Niizumi2023byola}.
We adopted a simple and uniform classification protocol across these datasets for all pre-trained models to ensure a fair and robust basis for comparative analysis.

The contributions of this paper are as follows.
Firstly, we provide valuable guidelines for the selection of pre-trained audio embeddings with regard to the influence of \ac{ts}. 
Notably, we highlight the efficacy of \ac{ast} models such as PaSST and BEATs, even when operating with smaller \ac{ts} across all tasks.
Reducing \ac{ts} substantially decreases memory and computational overhead at inference time. 
Secondly, through the selection of the optimal \ac{ts}, we improve the state-of-the-art results on the OpenMIC dataset using BEATs and PaSST without resorting to any fine-tuning procedure.
Finally, we explored the utility of intermediate layer representations when using PaSST and BEATs, enhancing overall results.

\vspace{-0.7em}

\section{Method}
\label{sec:method}

The aim is to use a frozen pre-trained model, denoted as $\f$, to project audio instances into embeddings. These embeddings are then used to make class predictions using a simple classification probe denoted as $\g$.
\vspace{-0.5em}
\subsection{Embedding projection using $\f$}
%\noindent\textbf{Embedding projection using $\f$} \
Given an audio instance  $\X=(x_n)_{n\in\{1,...,N\}}$, where $N$ represents the number of audio samples, we project $\X$ into an embedding space using the pre-trained  model $f(\cdot)$.
This result in an embedding sequence $\E=(\e_t)_{t\in\{1,...,T\}}$, where $T$ corresponds to the number of embedding vectors, and $\e_t \in \mathbb{R}^m$ with $m$ indicating the dimensionality of the embedding. 
Typically, generating a single embedding entails feeding a segment of $\X$ as input of $\f$.
We denote by \acl{ts}, $\ts$, the duration (in seconds) of this segment. 
The number of embeddings $T$ is a function of $N$, and the value of $\delta_{t}$. Precisely, $T$ can be calculated as follows: $T = \lfloor \frac{N}{\ts\times s_r } \rfloor$, where $s_r$ is the sample rate of $\X$. 
To illustrate, if we consider $\delta_t=1$ and an audio instance of 10 s, we obtain 10 embeddings $(\e_t)_{t\in\{1,...,10\}}$.
For a given embedding system, $\delta_t$ should not be chosen to be too small and is always assumed to be significantly greater than the length of the analysis window used to compute the input representation (typically a spectrogram).
\vspace{-0.5em}
\subsection{Label prediction using $\g$}
\begin{figure}[t]
  \centering
  \centerline{\includegraphics[width=0.8\columnwidth]{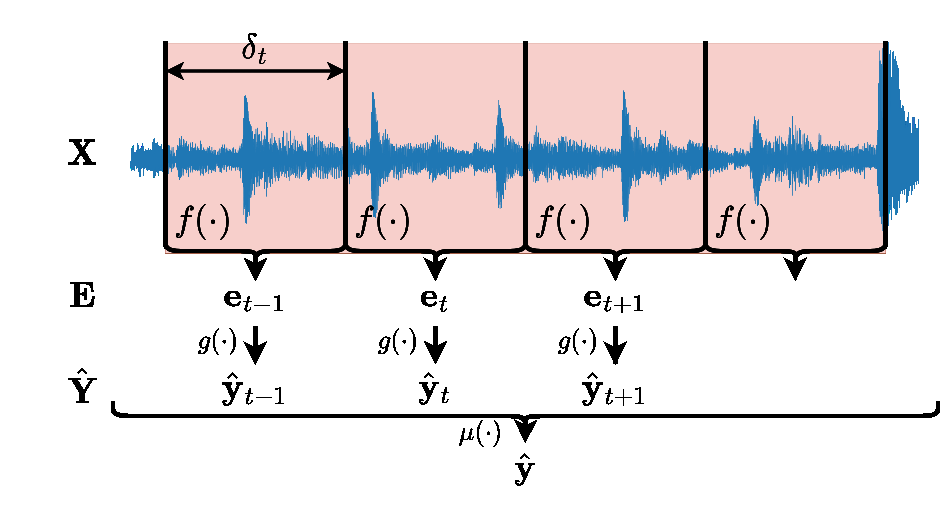}}
  \caption{Projection of an audio instance $\X$ into an embedding $\E$ with pre-trained model $\f$, using a Temporal Support $\ts$. This is followed by the label prediction facilitated by a classification probe $\g$ and aggregation operation $\mu(\cdot)$.}
  \label{fig:aggreg_tempo}
\end{figure}

%\noindent\textbf{Label prediction using $\g$} \ 
To predict labels from the embedding representation $\E$, we employ a classification probe $\g$ resulting in $\Yh = g(\E) = (\hat{\y}_t)_{t\in\{1,...,T\}}$. This yields a label prediction $\hat{\y}_t$ for each time step of the embedding representation.
Yet, we place ourselves in a so-called \textit{weakly labeled data} setting where for each audio instance $\X$ we only have a global label $\y \in \{0,1\}^L$ with $L$ representing the number of classes. Here, $y_l = 1$ indicates the presence of the class $l$ in the audio $\X$. 
Since the projection head yields a label prediction for each time step of $\E$, it is necessary to aggregate these ``local" predictions $\hat{\y}_t$ to get the final estimated label $\yh$, as illustrated in Figure \ref{fig:aggreg_tempo}. 
We consider two alternative aggregation methods: a learnable aggregation operation, commonly referred to as attention \cite{McFee18attention}, and simple average pooling, also known as mean temporal integration. We denote the aggregation function by $\mu(.)$, such that $\yh = \mu(g(\E)) \in [0,1]^L$, where $\yh$ is the final clip-level label prediction for an instance $\X$. Specifically, we refer to $\mu(.)$ as $\um$ for mean-based aggregation and $\ua$ for attention-based aggregation.

% New lay-out.
\begin{table*}
    \centering
    \footnotesize
    \label{tab:results}
    \begin{tabular}{lccccccccc}
        \toprule
        \multirow{2}{*}{Model} & \multirow{2}{*}{$\ts$} &\multicolumn{2}{c}{OpenMIC}  & \multicolumn{2}{c}{TAU Urban} & \multicolumn{2}{c}{ESC-50} & \multirow{2}{*}{Emb. Size} & \multirow{2}{*}{\#Param. $\f$}\\
        
        \cmidrule(lr){3-4} \cmidrule(lr){5-6} \cmidrule(lr){7-8}
        & & $\um$ & $\ua$ & $\um$ & $\ua$ & $\um$ & $\ua$ \\
        
        \midrule        
        BYOL-A v2 & \multirow{3}{*}{1} & 0.792 $\pm$ 0.001 & 0.797 $\pm$ 0.003 & 52.5 $\pm$ 1.4 & 50.6 $\pm$ 1.7 & 69.1 $\pm$ 1.4 & 68.7 $\pm$ 1.1 & 3072 & 6.3M\\
        PaSST &                     & 0.851 $\pm$ 0.001 & 0.860 $\pm$ 0.002 & 63.3 $\pm$ 0.4 & 62.0 $\pm$ 0.5 & 93.1 $\pm$ 0.2 & 93.0 $\pm$ 0.4 & 768 & 87M\\
        BEATs &                     & 0.852 $\pm$ 0.001 & 0.865 $\pm$ 0.001 & \textbf{67.5 $\pm$ 0.2} & 61.0 $\pm$ 4.3 & 93.2 $\pm$ 0.1 & 93.4 $\pm$ 0.4 &  $48 \cdot 768$ & 90M\\

        \midrule
        BYOL-A v2 & \multirow{3}{*}{3} & 0.805 $\pm$ 0.001 & 0.804 $\pm$ 0.005 & 53.9 $\pm$ 0.9 & 52.3 $\pm$ 0.9 & 71.2 $\pm$ 1.1 & 72.6 $\pm$ 1.0 & 3072 & 6.3M\\
        PaSST &                   & 0.866 $\pm$ 0.001 & 0.865 $\pm$ 0.000 & 65.0 $\pm$ 0.4 & 64.5 $\pm$ 0.5 & 95.7 $\pm$ 0.1 & 95.0 $\pm$ 0.1 & 768 & 87M\\
        BEATs &                   & 0.862 $\pm$ 0.000 & 0.866 $\pm$ 0.002 & 66.8 $\pm$ 0.2 & 64.9 $\pm$ 1.4 & 95.4 $\pm$ 0.1  & 93.4 $\pm$ 0.3 & $144 \cdot 768$ & 90M\\
        
        \midrule
        BYOL-A v2 & \multirow{3}{*}{5}  & 0.806 $\pm$ 0.002 & 0.808 $\pm$ 0.003 & 53.8 $\pm$ 1.1 & 53.6 $\pm$ 0.9 & 72.8 $\pm$ 1.8 & 74.0 $\pm$ 1.1 & 3072 & 6.3M\\
        PaSST &                         & 0.866 $\pm$ 0.001 & 0.868 $\pm$ 0.001 & 66.5 $\pm$ 0.5 & 65.9 $\pm$ 1.0 & \textbf{96.8 $\pm$ 0.2} & 96.6 $\pm$ 0.2 & 768 & 87M  \\
        BEATs &                         & \textbf{0.869 $\pm$ 0.002} & \textbf{0.869 $\pm$ 0.001} & \textbf{67.5 $\pm$ 0.2} & 65.4 $\pm$ 2.6 & 96.1 $\pm$ 0.0  & 95.7 $\pm$ 0.3  & $248 \cdot 768$ & 90M\\
        
        \midrule
        BYOL-A v2 & \multirow{3}{*}{10} & 0.803 $\pm$ 0.001 & 0.805 $\pm$ 0.002 & 52.4 $\pm$ 1.5 & 54.7 $\pm$ 0.8 & - & - & 3072 & 6.3M\\
        PaSST &                         & 0.861 $\pm$ 0.001 & 0.857 $\pm$ 0.001 & 66.7 $\pm$ 0.5 & 66.9 $\pm$ 0.4 & - & - & 768 & 87M  \\
        BEATs &                         & 0.866 $\pm$ 0.000 & 0.867 $\pm$ 0.000 & \textbf{67.5 $\pm$ 0.3} & 67.2 $\pm$ 1.1 & - & - & $496 \cdot 768$ & 90M\\
       
        \midrule
        \multicolumn{10}{c}{Results from papers}\\
        \midrule
        ResAtt \cite{Zhong23openmic} & 10 & \multicolumn{2}{c}{0.860} & \multicolumn{2}{c}{-} & \multicolumn{2}{c}{-} &  2048 & - \\
        PaSST-S \cite{Koutini22passt} &    10/5 & \multicolumn{2}{c}{0.843}           & \multicolumn{2}{c}{75.6} & \multicolumn{2}{c}{96.8} & 768 & 87M\\
        BEATs \text{iter3+}\cite{Chen22beats}  & 5 & \multicolumn{2}{c}{-} & \multicolumn{2}{c}{-} & \multicolumn{2}{c}{98.1} &  248 · 768 & 90M  \\
        
        \bottomrule
    \end{tabular}
    \caption{Results for frozen pre-trained embeddings with varying $\ts$ and two temporal aggregations $\um$ and $\ua$. Metrics include mAP for OpenMIC, and accuracy for TAU Urban and ESC-50.}
    \label{tab:results}
\end{table*}

\vspace{-0.5em}
\section{Pre-Trained Embeddings $\f$}
\label{sec:pretrainedemb}
 The selection of the following embeddings was guided by the criteria outlined in the introduction.

\noindent \textbf{BYOL-A} \cite{Niizumi2023byola} is a \ac{ssl} model that extends the BYOL \cite{Grillbyol} method tailored to learn representations that capture foreground acoustic events and sound texture details. Its encoder relies on a \ac{cnn} architecture trained on AudioSet. The training data consists of 1-second long audio segments, corresponding to $\ts=1$.

\noindent \textbf{PaSST} \cite{Koutini22passt}, falls under the\ac{ast}  category \cite{Gong21ast} and draws inspiration from the popular Vision Transformers (ViT) models \cite{TouvronVit}. Unlike BYOL-A, PaSST is trained in a supervised fashion on AudioSet. To reduce memory and computational costs, it employs patchout masking during training, enabling classification with incomplete patch sequences. The training segments in this case are 10 seconds long, $\ie$ $\ts=10$ s.

\noindent \textbf{BEATs} \cite{Chen22beats} similarly to PaSST, adopts an \ac{ast} architecture, but uses a \ac{ssl} iterative training procedure. Optimization involves a tokenizer model for generating pseudo-labels and knowledge distillation. BEATs is also trained on the AudioSet dataset using full-length audio instance, hence with $\ts=10$ s.

It is important to bear in mind that, regardless of the \ac{ts} that was used for training the embedding networks, one may always exploit them to extract embeddings with a different temporal support. To clarify this with an example: although BEATs was initially trained with $\ts=10$, we are interested in examining its behavior when, for a specific downstream classification task, we opt for a different temporal support  (\textit{e.g.}, $\delta_t = 3$) for extracting these embeddings. These embeddings will then serve as input features for training the classifier.

\vspace{-0.8em}
\section{Experimental Setup}
\label{sec:experimentalstudy}

\subsection{Datasets}
\label{sec:datasets}

We employ three datasets for our study: OpenMIC for musical instrument recognition, TAU Urban for audio scene classification, and ESC-50 for environmental sound classification, which are described in the following. 

\noindent \textbf{OpenMIC} \cite{Humphrey18openmic} is sourced from the Free Music Archive \cite{Defferrard17fma}, featuring a broad and diverse set of musical genres and instruments. This dataset comprises 20 possible instrument classes, with each of the 20,000 10-second audio clip annotated with at least one instrument label at the clip-level, making it suitable for multi-label classification.
%While the annotations are balanced per class (2 annotations per clip on average), the positive and negative annotations per clip are unbalanced. 
In our experiments, we use the provided training and testing splits and further partition the training set, keeping 15\% of the samples for validation purposes as in \cite{Gururani19, Koutini21rf}. To handle missing labels within OpenMIC, we compute the loss exclusively for the observed labels, following the approach outlined in \cite{Gururani19, Koutini21rf}. For evaluation, we report the instrument-wise \ac{map} scores, as done in prior works such as \cite{Koutini21rf, Koutini22passt, Zhong23openmic}.

\noindent \textbf{TAU Urban Acoustic Scenes 2020 Mobile} \cite{Mesaros2018taudata} This dataset comprises 10-second audio recordings captured in various urban environments from 12 European cities. It covers the following scenes: airport, shopping mall, metro station, pedestrian street, public square, street traffic, tram, bus, metro, and park. Designed for multi-class classification, it contains a total of 23,040 samples. We used the official training and testing splits and further divided the training set by reserving 30\% of the samples for validation following the approach used in \cite{Koutini22passt}. We use classification accuracy as a metric for this dataset. 

\noindent  \textbf{ESC-50} \cite{Piczak15esc50} is a dataset intended for multi-class audio classification of environmental sounds. It consists of 2,000 5-second audio clips categorized into 50 classes with 40 examples per class. These classes can be grouped into five broader categories: Animals, Natural soundscapes\textbackslash water sounds, Human\textbackslash non-speech sound, Interior\textbackslash Domestic sounds, and Exterior\textbackslash urban noises. The audio samples are sourced from the Freesound.org project. We use a 5-fold cross-validation setting as it is customary for this dataset \cite{turian2022hear}. We use classification accuracy as a metric for this dataset.

\subsection{Training Setup}

We applied the methodology described in Section \ref{sec:method} to each dataset employing different pre-trained models $\f$ and varying \ac{ts}. For $\g$, we used a linear layer to map the embedding to the number of classes of the task, followed by either $\um$ or $\ua$ for temporal aggregation of predictions. We selected $\ua$ as described in \cite{Gururani19} for OpenMIC, and we utilized the attention mechanism commonly described in the DCASE Challenge \cite{Ronchini22dcaseattention} for TAU Urban and ESC-50. For training the classification probe, we used the Adam optimizer, a learning rate of $1e^{-4}$ with a binary cross entropy loss for the \ac{ir} task, and a learning rate\footnote{We used a fixed learning rate for every classification probe as a strategic trade-off.  While it may not optimize performance for every embedding type, it strikes a balance that ensures fairness and simplifies comparisons.} of $1e^{-3}$ for \ac{asc} and \ac{esc}, both using cross entropy loss. We used a batch size of 128 for \ac{ir} and a batch size of 32 for the other tasks. While batch size differences were not intended, our preliminary experiments revealed no performance changes linked to this variation. We kept the model with the best validation loss. Each experiment was conducted 5 times, and we report the averaged results along with the standard deviation.
Regarding the pre-trained models, we used publicly available weights: BYOL-A version 2 for BYOL-A, PaSST-S for PaSST, and \textit{iter3+} weights for BEATs.

\section{Results}

\label{sec:results}

In this section, we conduct a performance evaluation of the selected embeddings. To  this end, we extract embeddings with audio segments of varying durations: $\ts=1$ s, $\ts=3$ s, $\ts=5$ s and $\ts=10$ s, denoted by $\delta_1$, $\delta_3$, $\delta_5$ and $\delta_{10}$, respectively. The comprehensive results of this evaluation are presented in Table \ref{tab:results}.

% \begin{figure*}[!h]
%   \centering
%   \begin{subfigure}[b]{0.3\textwidth}
%          \centering
%          \includegraphics[width=\textwidth]{img/beats_vs_passt_openmic.png}
%          \caption{OpenMIC}
%          \label{fig:OpenMIC}
%      \end{subfigure}
%      \hfill
%     \begin{subfigure}[b]{0.3\textwidth}
%          \centering
%          \includegraphics[width=\textwidth]{img/beats_vs_passt_TAU.png}
%          \caption{TAU Urban}
%          \label{fig:TAUUrban}
%      \end{subfigure}
%      \hfill
%      \begin{subfigure}[b]{0.2855\textwidth}
%          \centering
%          \includegraphics[width=\textwidth]{img/beats_vs_passt_ESC.png}
%          \caption{ESC-50}
%          \label{fig:ESC50}
%      \end{subfigure}
%      \hfill
%   \caption{Influence between $\tstr$ and $\tsev$ on OpenMIC, TAU Urban and ESC-50}
%   \label{fig:single_ts}
% \end{figure*}

%\noindent \textbf{Influence of the \acl{ts}.} \ 
\subsection{Influence of the \acl{ts}}

First, it is evident that pre-trained \ac{ast} models consistently outperform BYOL-A across all $\ts$ configurations, which is to be expected due to the inherent differences in the number of parameters between these models. A more interesting finding is that pre-trained models are not necessarily more effective when used with the largest possible $\ts$.
For instance, all models achieve the best performance on the OpenMIC dataset for $\delta_5$. 
BEATs behaves similarly on the TAU Urban dataset, since results for $\delta_1$, $\delta_5$ and $\delta_{10}$ yielded the same results.
While BYOL-A achieves its highest score for $\delta_3$ on TAU Urban, PaSST achieves its best scores when using either $\delta_5$ or $\delta_{10}$.
Conversely, for ESC-50, all pre-trained models achieve the best results for the largest possible temporal support, which is $\delta_5$.
Overall, this highlights the importance of selecting the optimal $\ts$ value for each pre-trained model,  depending on the specific task, in order to attain the best possible performance.
This is of paramount importance for pre-trained \ac{ast} models, given the computational cost associated with larger \ac{ts}.
Indeed, let $L$ denote the sequence length of an attention layer within an \ac{ast} model. It's worth noting that the computation complexity grows quadratically, specifically in $\mathcal{O}(L^2)$ as the sequence length $L$ increases.
Consequently, opting for a smaller \ac{ts} implies processing sequences of much smaller length in each attention layer, which reduces drastically the computational burden during inference.
For instance, one should extract embeddings with $\delta_1$ rather than $\delta_{10}$ with BEATs on TAU Urban due to the significantly reduced computational inference cost while maintaining equivalent performance. %since for the same performance, the computational cost of inference is much smaller with $\delta_1$ compared to $\delta_{10}$.
Furthermore, it is noteworthy to highlight that we improve the state-of-the-art performance on OpenMIC with BEATs and PaSST for every value of $\ts$ (with the exception of $\delta_{10}$ for PaSST) without any fine-tuning effort. 
% The reason why the models perform better for some $\ts$ than for others remains to be explained and demonstrated.

\noindent\textbf{Discussion.} \ 
When it comes to \ac{ast} models, adjusting their \ac{ts} can be likened to altering the receptive field of a \ac{cnn}. Indeed, in \ac{cnn}s the receptive field is defined as the input area that contributes to the resulting embedding at a given position. In contrast, \ac{ast} models rely on the attention mechanism, where the entire input contributes to the final embedding (especially on short sequences as it is the case here), effectively determined by the input's length, $\ie \ts$.
Remarkably, prior research \cite{Koutini21rf} demonstrated that regularizing the receptive field of a \ac{cnn} for a specific task can significantly improve performance. Therefore, selecting the optimal \ac{ts} for pre-trained \ac{ast} models may be as advantageous as fine-tuning the receptive field of a \ac{cnn} model for a given task.

% However, for \ac{ast} models, modifying their \ac{ts} can be seen as equivalent to modifying the receptive field of a \ac{cnn}. 
% Indeed, in \ac{cnn}s the receptive field is defined as the area in the input that contributes to the corresponding embedding at a given position. But for \ac{ast}, the whole input determines the final embedding thanks to the attention mechanism, therefore the length of the input, $\ie \ts$.
% Yet, previous work \cite{Koutini21rf} has shown that regularizing the receptive field of a \ac{cnn} on a given task can considerably improve performances.
% Picking the optimal \ac{ts} for pre-trained \ac{ast} models could than be as beneficial as tuning the receptive field of a \ac{cnn} model on a given task.
% % However, this study did not reveal any general behavior on the scores across all datasets regarding the \ac{ts}.

\begin{table}[b]
    \centering
    \footnotesize
    \begin{tabular}{lcccc}
        \toprule
         \textbf{Model} & $\ts$ & OpenMIC & TAU Urban & ESC-50\\
 
        \midrule   

         \multirow{4}{*}{PaSST} & 1 & 0.860 $\pm$ 0.001 &             63.0 $\pm$ 0.4 & 93.3 $\pm$ 0.2 \\
         &                                     3 & 0.868 $\pm$ 0.001 & 65.2 $\pm$ 0.4 & 96.2 $\pm$ 0.0\\
         &                                     5 & 0.869 $\pm$ 0.002 & 67.1 $\pm$ 0.6 & \textbf{97.0 $\pm$ 0.0}\\
         &                                    10 & 0.868 $\pm$ 0.002 & 67.9 $\pm$ 0.4 & - \\
         
         \midrule 
         
         \multirow{3}{*}{BEATs} & 1 & 0.864 $\pm$ 0.002                       & 67.2 $\pm$ 0.2 & 93.6 $\pm$ 0.1 \\
         &                                     3 &  0.867 $\pm$ 0.001         & 67.4 $\pm$ 0.3 & 95.4 $\pm$ 0.1 \\
         &                                     5 & \textbf{0.870 $\pm$ 0.002} & \textbf{68.5 $\pm$ 0.3} & 96.0 $\pm$ 0.1 \\
         &                                    10 & \textbf{0.870 $\pm$ 0.002} & 68.2 $\pm$ 0.3 & - \\
         
        \bottomrule
    \end{tabular}
    \caption{Results for PaSST and BEATs using a learned weighted sum of the output of all layers.}
    
    \label{tab:layers}
\end{table}

%\noindent \textbf{Aggregation method.} \ 

\subsection{Aggregation method}
Our analysis of Table \ref{tab:results} reveals two distinct trends regarding the performance of the aggregation function $\mu(.)$. 
For the \ac{ir} task, consistent with \cite{Gururani19}, we observe that $\ua$ is more effective, particularly for small $\ts$ values, which results in an increased number of time steps in $\Yh$. 
In contrast, for \ac{asc} and \ac{esc} tasks, $\um$ aggregation has the ascendancy in most cases. There are few exceptions, notably BYOL-A with $\delta_{10}$ on TAU Urban, and $\delta_3$ and $\delta_5$ on ESC-50.
This is an interesting finding, as this aggregation approach is far more efficient, which was not expected. Indeed, given that the dynamics of target sound classes vary (regardless of the dataset), with some class activations being localized in time while other classes being active for longer periods, we expected to see $\ua$ be the best aggregation overall. 
Even if we would need to further investigate the causes that can explain this behavior, we believe that in the case of the multi-class nature of the problems, where events overlap in time, yet the model yields only one class, the training of the attention mechanism may be hindered, especially in this regime of rather small datasets.
%the type of classification (multi-class or multi-label) can be at the origin of the explanation.
\vspace{-0.5em}
%\noindent \textbf{Comparing results to previous works.} \ 
\subsection{Comparing results to previous works}

With equal \ac{ts}, $\delta_{10}$, and without fine-tuning, we achieve a higher score on OpenMIC with PaSST compared to \cite{Koutini22passt}.
This shows that fine-tuning large pre-trained models on smaller datasets may not always be beneficial. This phenomenon is akin to \textit{forgetting} which may limit the generalization ability of the fine-tuned variant compared to the original one. On the other hand, on the TAU Urban dataset, we obtain scores with PaSST that are below those obtained in \cite{Koutini22passt}, where fine-tuning is used.
For BYOL-A, our results on ESC-50 deviate from those reported in \cite{Niizumi2023byola} and \cite{lWang2022hares}. This divergence can be partly attributed to differences in methodology. In \cite{Niizumi2023byola}, the authors normalize the embeddings before training the probe $\g$ and optimize the learning rate to maximize the scores. 
In \cite{lWang2022hares}, the authors employed a meticulously tuned support vector machine classifier in conjunction with data augmentation, which is expected to be more effective than the shallow probe we use across all tasks. Such improvements and careful optimizations are beyond the scope of this study whose sole focus is on the influence of the temporal support with a neutral, lightweight downstream classification strategy kept fixed across all tasks.

\vspace{-0.5em}

\subsection{Exploiting outputs of all embedding layers}
%\noindent\textbf{Exploiting outputs of all embedding layers.} \ 
This last experiment looks into exploiting the outputs of all intermediate layers of the embedding network as opposed to solely relying on the last one. Indeed, different types of information are encoded across these layers which may be beneficial to the classification task. This practice aligns with common conventions in the Speech \ac{ssl} domain, where benchmarking pre-trained embedding models often involves leveraging multiple layers.  In particular, the SUPERB benchmark \cite{Yang2021superb} suggests using a weighted sum of the outputs from each layer as the final embedding to be exploited by the downstream classifier $\g$, using $z = \sum_{l=1}^{12} \alpha_l z^{(l)}$, where $z^{(l)}$ is the output of layer $l$. The weights $\alpha_l$ are trained jointly with the downstream classifier, while the representations $z^{(l)}$ are merely the intermediate outputs of the embedding network without fine-tuning. This is because fine-tuning such networks is costly and often sub-optimal, with catastrophic forgetting easily compromising performance, as explained earlier.

Hence, we explore the impact of combining the outputs from all layers of the pre-trained \ac{ast} models, PaSST and BEATs, across the three datasets for different \ac{ts}.
This yields a total of 11 intermediate representations along with the final layer embedding that are used (as a weighted combination) as the final embedding for both PaSST and BEATs. We select the most effective temporal aggregation method for each dataset, namely $\ua$ for OpenMIC and $\um$ for TAU Urban and ESC-50.
The results of this analysis are presented in Table \ref{tab:layers}. 
In general, we observe only marginal performance improvements when leveraging outputs from all layers considering the augmentation of the dimensionality of the used features, with no noticeable influence of \ac{ts} on the results. 
Nevertheless, this approach enables us to improve the state-of-the-art mAP score on OpenMIC, achieving a remarkable 0.87, and outperforming \cite{Koutini22passt} on ESC-50  without the need of fine-tuning the pre-trained model. 

\section{Conclusion}
\label{sec:conclusion}

In this study, we have conducted a comprehensive experimental analysis to investigate the influence of \acl{ts} on audio embeddings extracted with BYOL-A, PaSST, and BEATs, across tasks including \acl{ir}, \acl{asc}, and \acl{esc}.
Our findings underscore the crucial importance of selecting the optimal \acl{ts} based on both the pre-trained model and the task at hand. 
This is particularly compelling as operating \acl{ast} pre-trained models with smaller \acl{ts} significantly reduces their inference computational cost.
With adequate \acl{ts}, these pre-trained models obtain competitive results across all tasks, all without the need for fine-tuning. Moreover, we improve the state-of-the-art result on the OpenMIC dataset.
Notably, by using a weighted combination encompassing all layer outputs from PaSST and BEATs, we further improved over our prior results.
Future works could delve into understanding how the weights of the last layers correlate with the \acl{ts} as well as explore more pre-trained models and tasks.

\section{Acknowledgement}
This work was supported by the Audible project, funded by French BPI. This work was performed using HPC resources
from GENCI-IDRIS.

\vspace{-0.5em}
% References should be produced using the bibtex program from suitable
% BiBTeX files (here: strings, refs, manuals). The IEEEbib.bst bibliography
% style file from IEEE produces unsorted bibliography list.
% -------------------------------------------------------------------------
\bibliographystyle{IEEEbib}
\bibliography{refs}

\end{document}